\documentclass[twocolumn, a4paper,14pt]{revtex4}
\usepackage{graphicx}
\usepackage{amsmath,amsthm,amssymb,amscd}
\usepackage{latexsym}
\usepackage{indentfirst}
\usepackage{subfigure}
\usepackage[top=0.5in,bottom=0.5in,left=0.5in,right=0.5in]{geometry}

\begin{document}

\title{
  \bf Comment on \lq\lq Fluctuation Theorem Uncertainty Relation"
   and \lq\lq Thermodynamic Uncertainty Relations from Exchange Fluctuation Theorems"}

\author{Yunxin Zhang}%\thanks{xyz@fudan.edu.cn}%\email{xyz@fudan.edu.cn}
\affiliation{Shanghai Key Laboratory for Contemporary Applied Mathematics,
 School of Mathematical Sciences, Fudan University, Shanghai 200433, China.
}

\begin{abstract}
In recent letter [Phys.~Rev.~Lett {\bf 123}, 110602 (2019)], Y.~Hasegawa and T.~V.~Vu derived a thermodynamic uncertainty relation. But the bound of their relation is loose. In this comment, through minor changes, an improved bound is obtained. This improved bound is the same as the one obtained in [Phys.~Rev.~Lett {\bf 123}, 090604 (2019)] by A.~M.~Timpanaro {\it et. al.}, but the derivation here is straightforward.
\end{abstract}

\maketitle

Let $P(\sigma,\phi)$ be the probability that we observe the total entropy production $\sigma$ and the observable $\phi$ in the forward process, and assuming the strong detailed fluctuation theorem holds, {\it i.e.}, $P(\sigma,\phi)/P(-\sigma,-\phi)=e^{\sigma}$. In \cite{Hasegawa2019}, it is obtained that
$$
\begin{aligned}
&\langle\phi\rangle=\left\langle\phi\tanh\left({\sigma}/{2}\right)\right\rangle_Q,\quad
\langle\phi^2\rangle=\left\langle\phi^2\right\rangle_Q,\cr
&\langle\sigma\rangle=\left\langle\sigma\tanh\left({\sigma}/{2}\right)\right\rangle_Q,\cr
\end{aligned}
$$
where
$$
\begin{aligned}
&\langle\alpha(\sigma,\phi)\rangle:=\int_{-\infty}^{\infty}d\sigma\int_{-\infty}^{\infty}d\phi P(\sigma,\phi)\alpha(\sigma,\phi),\\
&\langle\alpha(\sigma,\phi)\rangle_Q:=\int_{0}^{\infty}d\sigma\int_{-\infty}^{\infty}d\phi Q(\sigma,\phi)\alpha(\sigma,\phi),
\end{aligned}
$$
with $Q(\sigma,\phi):=(1+e^{-\sigma})P(\sigma,\phi)$.
By Cauchy-Schwarz inequality,
\begin{eqnarray}\label{eq1}
\langle\phi\rangle^2\le \left\langle\phi^2\right\rangle_Q
\left\langle\tanh^2\left({\sigma}/{2}\right)\right\rangle_Q
=\left\langle\phi^2\right\rangle
\left\langle\tanh^2\left({\sigma}/{2}\right)\right\rangle_Q.
\end{eqnarray}

\begin{figure}
  \includegraphics[width=9cm]{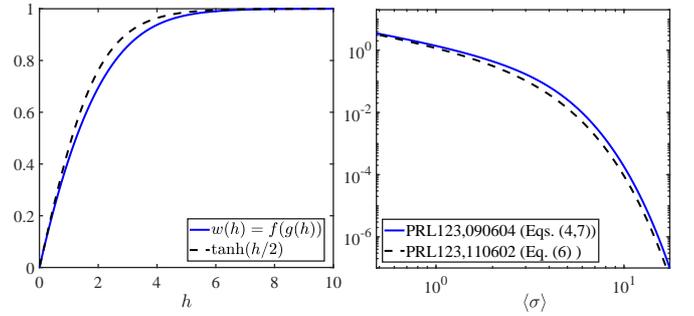}\\
  \caption{({\bf Left}) Figures of function $w(h)=f[g(h)]=\tanh^2\left(g(h)/2\right)$ and $\tan(h/2)$. ({\bf Right}) Bounds of the thermodynamic uncertainty relation obtained in \cite{Timpanaro2019} (see Eq.~(\ref{eq4})) and \cite{Hasegawa2019} (see Eq.~(\ref{eq5-1})).}\label{Fig1}
\end{figure}

For convenience, denote
$$
f(\sigma):=\tanh^2\left({\sigma}/{2}\right), \quad h(\sigma):=\sigma\tanh\left({\sigma}/{2}\right).
$$
Obviously, $\langle h(\sigma)\rangle_Q=\langle \sigma\rangle$ and $h(\sigma)$ is a monotonically strictly increasing function for $\sigma\ge0$, see \cite{Merhav2010}. We denote the inverse function of $h$ by $g$, {\it i.e.}, $g\textbf{(}h(\sigma)\textbf{)}=\sigma$. It can be mathematically shown that $w(h):=f[g(h)]$ satisfies $w'(h)>0$ and $w''(h)<0$ for $h>0$, which means $w(h)$ is a concave function, see Fig.~\ref{Fig1}. By Jensen inequality,
\begin{eqnarray}\label{eq2}
\left\langle\tanh^2\left({\sigma}/{2}\right)\right\rangle_Q&=&
\left\langle f(\sigma)\right\rangle_Q=
\left\langle f[g\textbf{(}h(\sigma)\textbf{)}]\right\rangle_Q\cr
&\le& f[g\textbf{(}\left\langle h(\sigma)\right\rangle_Q\textbf{)}]
=f[g\textbf{(}\langle \sigma\rangle\textbf{)}].
\end{eqnarray}
From Eqs.~(\ref{eq1},\ref{eq2}), we obtain
\begin{eqnarray}\label{eq3}
\langle\phi\rangle^2\le \left\langle\phi^2\right\rangle f[g\textbf{(}\langle \sigma\rangle\textbf{)}].
\end{eqnarray}
Therefore,
\begin{eqnarray}\label{eq4}
\frac{\textrm{Var}[\phi]}{\langle\phi\rangle^2}\ge\frac{1}{f[g\textbf{(}\langle \sigma\rangle\textbf{)}]}-1=\textrm{csch}^2\left(\frac{g\textbf{(}\langle \sigma\rangle\textbf{)}}{2}\right).
\end{eqnarray}

Since $f[g\textbf{(}\langle \sigma\rangle\textbf{)}]=\tanh^2(g\textbf{(}\langle \sigma\rangle\textbf{)}/2)<\tanh(\langle \sigma\rangle/2)$, the bound given in Eq.~(\ref{eq3}) is stricter than the one given in \cite{Hasegawa2019}, which is \begin{eqnarray}\label{eq5}
\langle\phi\rangle^2\le \left\langle\phi^2\right\rangle \tanh(\langle \sigma\rangle/2),
\end{eqnarray}
see Fig.~\ref{Fig1}. Consequently, the thermodynamic uncertainty relation given in Eq.~(\ref{eq4}) is stricter than the one given in \cite{Hasegawa2019}, which is \begin{eqnarray}\label{eq5-1}
\frac{\textrm{Var}[\phi]}{\langle\phi\rangle^2}\ge \coth\left(\frac{\langle \sigma\rangle}{2}\right)-1=\frac{2}{e^{\langle \sigma\rangle}-1}.
\end{eqnarray}

Finally, the bound given in Eq.~(\ref{eq4}) is the same as the one given in \cite{Timpanaro2019}, which is
\begin{eqnarray}\label{eq6}
\frac{\textrm{Var}[\phi]}{\langle\phi\rangle^2}\ge \textrm{csch}^2\left(\hat{g}\left (\frac{\langle \sigma\rangle}{2}\right)\right),
\end{eqnarray}
with $\hat{g}$ the inverse function of $\sigma\tanh \sigma$. Actually, it can be easily shown that $g\textbf{(}\langle \sigma\rangle\textbf{)}/2=\hat{g}\textbf{(}\langle \sigma\rangle/2\textbf{)}$. So, $\textrm{csch}^2(g\textbf{(}\langle \sigma\rangle\textbf{)}/2)=\textrm{csch}^2(\hat{g}\textbf{(}\langle \sigma\rangle/2\textbf{)})$, and the thermodynamic uncertainty relation given in Eq.~({\ref{eq4}}) is the same as the one obtained in \cite{Timpanaro2019} (Eq. (\ref{eq6})), but the derivation here is more straightforward.

\end{document}